\documentclass[11pt,twoside]{article}
\usepackage{CAGN2019}
\usepackage{graphicx} 
\usepackage{amsmath}
\usepackage[T1]{fontenc} 

\usepackage{latexsym}
\usepackage{verbatim}

\usepackage{ifpdf}  
\ifpdf  
      \DeclareGraphicsExtensions{.pdf,.png,.jpg}  
\else  
      \DeclareGraphicsExtensions{.eps}  
\fi 

\def \HII{{H{\sc ii}}}
\def \ldo{$\lambda$}
\def \ldos{$\lambda\lambda$}
\newcommand{\TeO}[1]{$T_{\rm{e}}$([O{#1}])}      
\newcommand{\TeS}[1]{$T_{\rm{e}}$([S{#1}])}

\begin{document}

\vskip 1.0cm
\markboth{A. I. Díaz}{Using Sulfur as nebular abundance tracer}
\pagestyle{myheadings}
%
%
\vspace*{0.5cm}
\parindent 0pt{Invited Review}


\vspace*{0.5cm}
\title{Using Sulfur as metallicity tracer in galaxies}

\author{A. I. ~Díaz$^{1,2}$}
\affil{$^1$ Department of Theoretical Physics, Universidad Autónoma de Madrid, Spain and 
$^2$ CIAFF, Universidad Autónoma de Madrid, Spain}

\begin{abstract}
We review the present methodology for the use of Sulfur as global metallicity tracer in galaxies, which allows performing a complete abundance analysis using mainly the red to near infrared spectral region, and extending the range of directly derived abundances up to 5 times the S solar photospheric value. The empirical calibration of Sulfur via the S$_{23}$ parameter is also reviewed. 

\bigskip
 \textbf{Key words: } galaxies: abundances --- galaxies: ISM --- techniques: imaging spectroscopy

\end{abstract}

\section{Introduction}

Oxygen is the most abundant element after Hydrogen and Helium. It is produced in high mass stars and hence it is returned to the ISM in a short time scale. Therefore, when measured in gaseous nebulae, its abundance can be associated with “present day abundances”. In the case of nebulae ionized by young, hot stars, given the range of their effective temperatures, Oxygen is present in the form of O$^+$ and O$^{++}$. When these ions collide with free electrons they get excited to higher energy levels and subsequently decay through radiative transitions producing forbidden emission lines of [OII] (\ldos{3727,29} \AA ) and [OIII] (\ldos{4959,5507} \AA ), readily accessible in the optical spectral region. Therefore the metallicity of gaseous nebulae has traditionally been traced by Oxygen. However, their use is restricted to abundances lower than solar since these lines act as very effective coolants and, at high abundances, nebulae become too cool for the lines to be detected.  So, what about using Sulfur? Sulfur is also produced in massive stars and its yield is supposed to follow that of Oxygen. Although less abundant than Oxygen, in principle, sulfur can also be used as abundance tracer providing similar information and, since the [SII] (\ldos{6716, 6731} \AA ) and [SIII] (\ldos{9069,9532} \AA ) lines take place at the red spectral region, they present several interesting observational advantages: they are less contaminated by stellar absorption lines from metals (there are many more metal lines in the blue) and less affected by reddening.  The lines can be measured relative to nearby Hydrogen recombination lines, thus minimizing the effects of reddening and flux calibration. Furthermore, there is almost no depletion of Sulfur onto dust grains in the diffuse ISM. But, most importantly: the [SIII] lines are clearly detected at high abundances where the “strong” [OIII] lines are very weak or almost undetectable due to the low electron temperature.  

The two samples selected for study consist of nebulae ionized by young massive stars located in the disks of spiral galaxies  and dwarf galaxies dominated by star formation bursts, henceforth  called DHR and \HII Gal respectively. High quality spectroscopic data have been taken from the literature and cover at least, the spectral range from \ldo{3700} \AA\ to \ldo{9600}  \AA\ at moderate to high resolution. In all cases the [SIII] line at \ldo{6312} \AA\ has been measured and, in combination with the strong nebular lines at \ldos{9069,9532} \AA , allows the derivation of the electron temperature \TeS{III}. 

\begin{figure}  
\begin{center}
\includegraphics[angle=0,height=5.0cm]{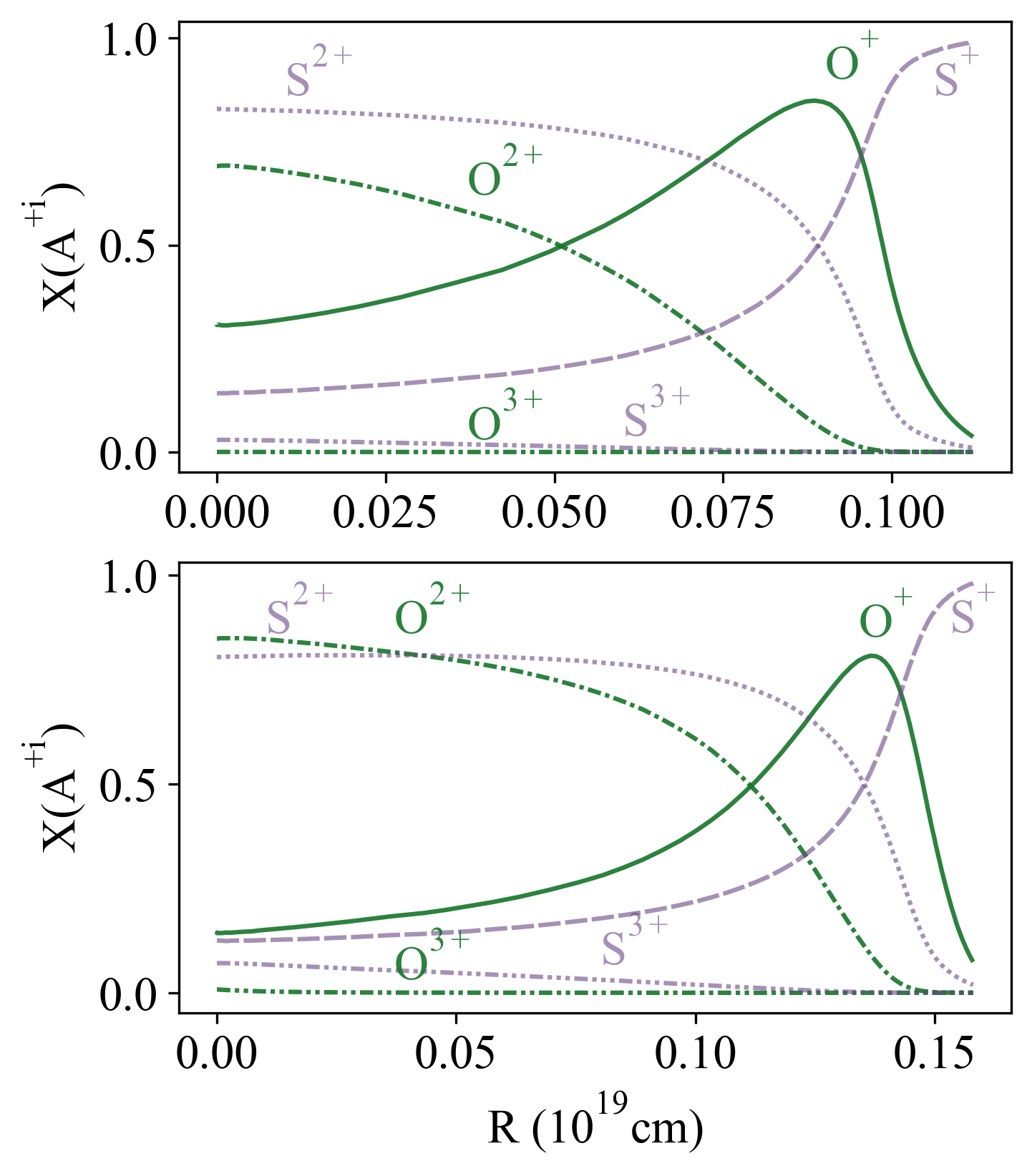}
\hspace*{0.3cm}
\includegraphics[angle=0,height=5.0cm]{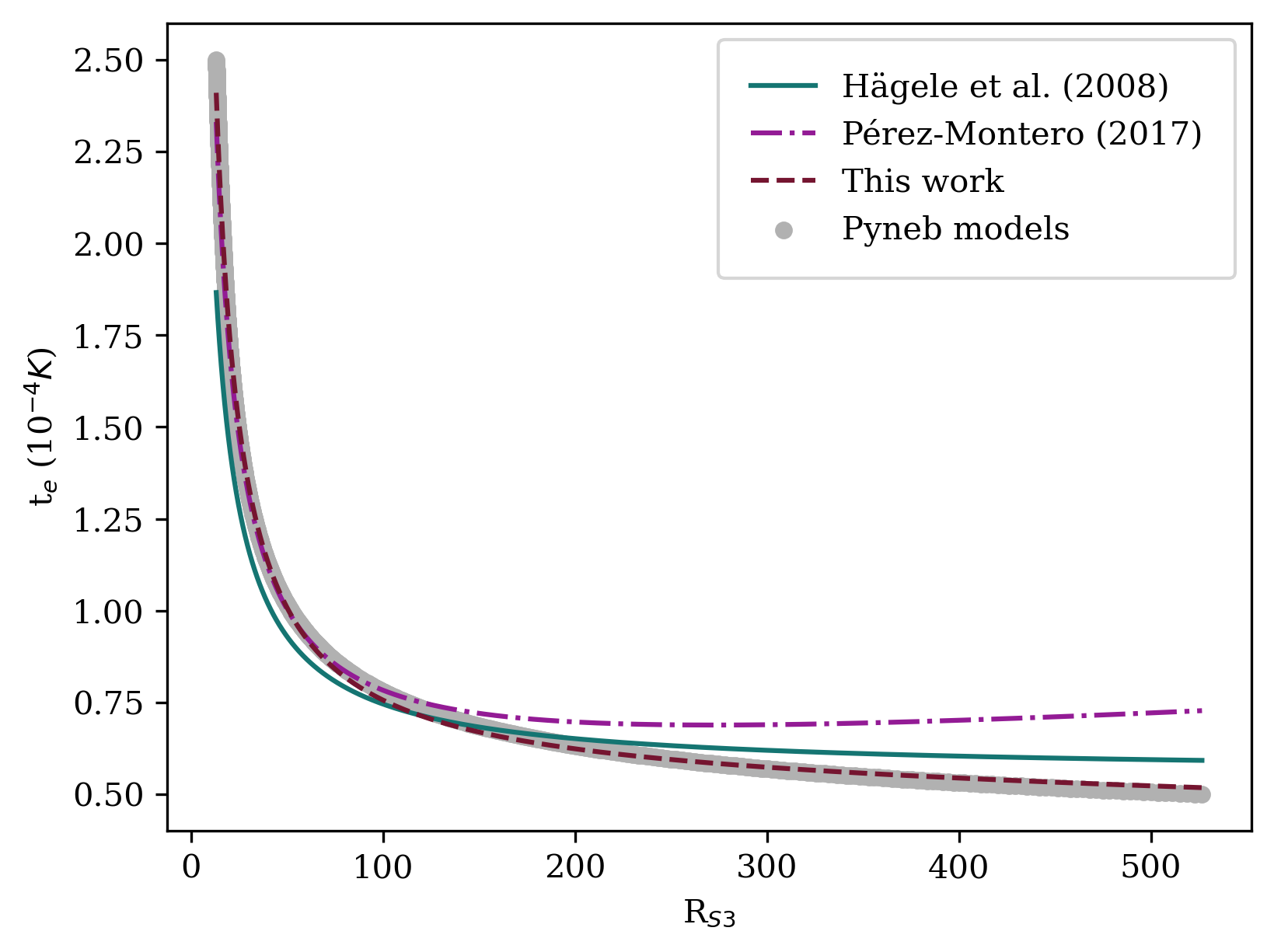}
\end{center}
\caption{{\it Left:} Ionization structure of a Cloudy model of low density (100 cm$^{-3}$) \HII\ regions of half solar abundance ionized by a 5 Ma star cluster, representative of the objects in the DHR sample (lower panel) and \HII regions of low metallicity (Z=0.004) ionized by a young (1 Ma) star cluster of 10$^5$ M$_{\odot}$, representative of the objects in sample \HII Gal (lower panel.
{\it Right:} The Sulfur temperature, $T_e$([SIII]) as a function of the line ratio $R_{S3}$, defined in the text.}
 \label{Diaz-fig1}
\end{figure} 

\section{Direct abundance derivation: Oxygen vs Sulfur}

The intensities of forbidden lines, as the ones we are talking about, depend exponentially of the electron temperature, ${T}_{{\rm{e}}}$. Different ions originate in regions with different ${T}_{{\rm{e}}}$ and a certain ionization structure for the nebula has to be assumed. 
In the case of Oxygen, \TeO{III} is derived from the ratio of auroral to nebular lines at \ldo{4363}\AA\ and \ldos{4959,5007}\AA .
Most of the O is in the form of O$^+$ and O$^{2+}$ and their relative contributions to the total O/H abundance depends on the
degree of ionization of the nebula. Only for exceptionally high excitation objects, those showing He{\sc ii} emission at \ldo{4686} \AA , a small contribution by O$^{3+}$ might be present. In the case of Sulfur, \TeS{III} is derived from the ratio of auroral to nebular lines at \ldo{6312} and \ldos{9069,9532} \AA . Most of the S is also in the form of S$^+$ and S$^{2+}$, but in most cases S$^{2+}$ is the dominant specie. However, a certain contribution by S$^{3+}$ is expected in high excitation objects for which ionization correction factors (ICF) have to be derived.

Regarding the ionization structure, a three-zone approximation is usually assumed for the derivation of the Oxygen abundance, where O$^{2+}$  and O$^+$, S$^+$ are produced in high and low ionization regions respectively, characterized by temperatures \TeO{III} and \TeO{II}; S$^{2+}$ is assumed to be produced in an intermediate excitation zone characterized by \TeS{III}. In the case of Sulfur, Figure \ref{Diaz-fig1} (left panel) shows the ionization structure computed from photo-ionization models for two cases representative of the objects in each of the studied samples.  In the two cases, both S$^+$ and S$^{2+}$ ions can be seen to to overlap in a transition zone that encompasses almost the whole nebula. According to this, we have assumed only two zones: S$^+$, S$^{2+}$, O$^+$ are formed in the same region, characterized by \TeS{III} while S$^{3+}$, O$^{2+}$ are formed in a higher ionization zone characterized by \TeO{III}.

Following this description we have derived the S$^+$/H$^+$ and S$^{2+}$/H$^+$ ionic abundances of 256 independent observations of disc \HII\ regions (including regions in The Milky Way Galaxy, labelled MW, and the Magellanic Clouds, labelled MC) and 95 independent observations of \HII\ galaxies, mostly compact, some of them with several star-forming knots. We have also derived Oxygen abundances for those regions with data on the [OIII] \ldo{4363} \AA\ line, allowing to derive \TeO{III}. The software package PyNeb \citep{2015A&A...573A..42L} has been used with the atomic coefficients listed in \citet{2018MNRAS.478.5301F}.

\begin{figure}  
\begin{center}
\includegraphics[angle=0,height=4.0cm]{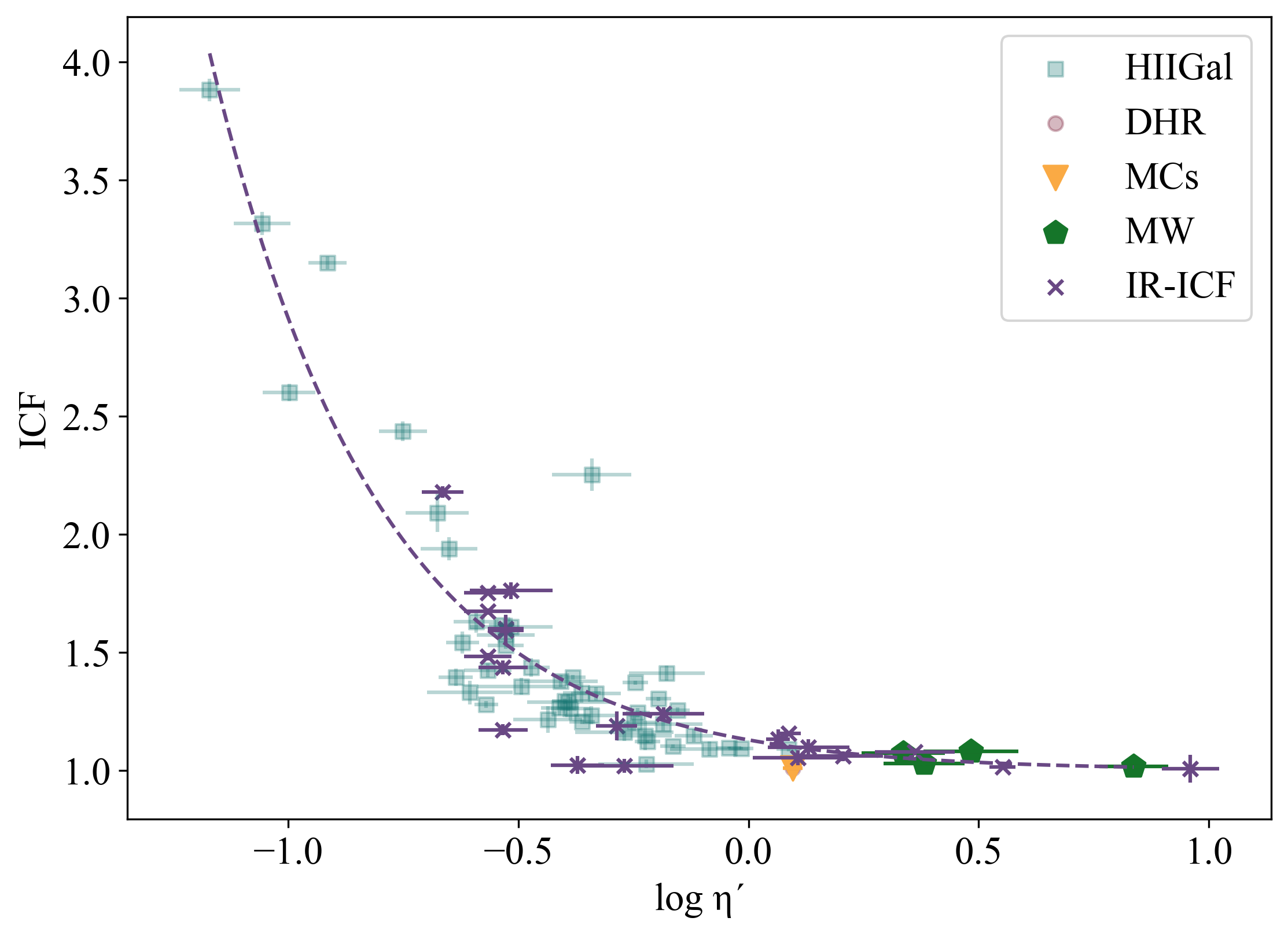}
\hspace*{0.3cm}
\includegraphics[angle=0,height=4.0cm]{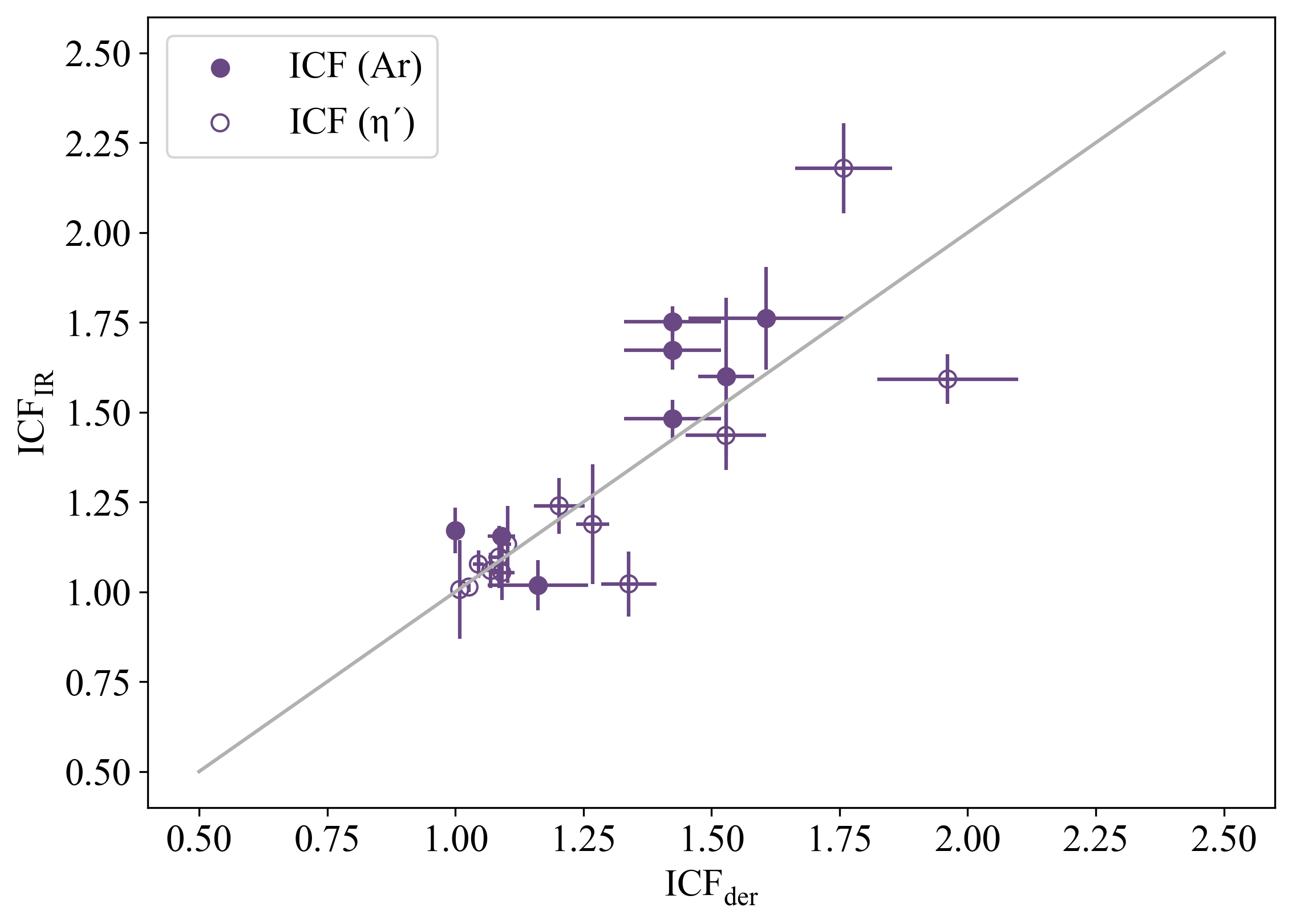}
\end{center}
\caption{{\it Left:} Relation between the derived ICF and the $\eta$' parameter. As can be seen, the ICF has a value $\simeq$ 1 for $\eta$'> 1.0.
{\it Right:} Comparison of the IR derived values of the Sulfur ICF with the ones calculated in this work. For objects labelled as ICF(Ar) they have been calculated from the derived Ar$^{2+}$/Ar$^{3+}$ ratio and  for objects labelled ICF($\eta$) from IR data.}
\label{Diaz-fig2}
\end{figure}

The S$^{2+}$ electron temperature, \TeS{III} has been derived from the ratio of the nebular to auroral [SIII] line intensities ($R_{S3}= I(\lambda 9069{\AA}+\lambda 9532{\AA})/I(\lambda 6312{\AA})$). We have used PyNeb to calculate $T_e([SIII])$ extending the temperature range down to 5500 K, since most expressions quoted in the literature do not apply to electron temperatures lower than about 8000 K, too high for many disk \HII\ regions present in our sample (see Figure \ref{Diaz-fig2}, right panel).

Total S/H abundances are found as: 

\begin{equation}
S/H = ICF \times \frac{S^{+}+S^{2+}}{H^+}
\end{equation}

\noindent where ICF is the ionization correction factor for Sulfur. Several schemes have been proposed for its derivation \citep[see for example][]{2016MNRAS.456.4407D}, most of them based on the ionization potential of Oxygen ions as compared to those of Sulfur. Our own proposal involves the relation found from photo-ionization models between the ionic ratios of Ar and S in contiguous ionization stages: $Ar^{+2}/Ar^{+3}$ vs. $S^{2+}/S^{3+}$ \citep{2018MNRAS.478.5301F} . The application of the procedure requires the measurement of the [ArIII] and [ArIV] emission lines at \ldo{7135} and \ldo{4740} \AA\ respectively.  

The ICF shows a certain dependence on the hardness of the ionizing radiation field which can be explored using the $\eta$' parameter defined by \citet{1988MNRAS.231..257V} as: 

\[ \eta^{'} = \frac{[OII]/[OIII]}{[SII]/[SIII]} \]
  
The relation between the derived ICF and $\eta$' is shown in Figure \ref{Diaz-fig2} (left panel). For objects without data on the [ArIV] \ldo{4740} \AA\ line we have used this relation to estimate the ICF. For some objects of the sample, there are data on the IR [SIV] line at 10.8 $\mu$m. For those objects we have derived the ICF directly and they compare well with our estimates \ref{Diaz-fig2} (right panel).

\section{Results and discussion}

\subsection{Sample characteristics}
\label{Sample characteristics}

\begin{figure}  
\begin{center}
\includegraphics[angle=0,height=6.0cm]{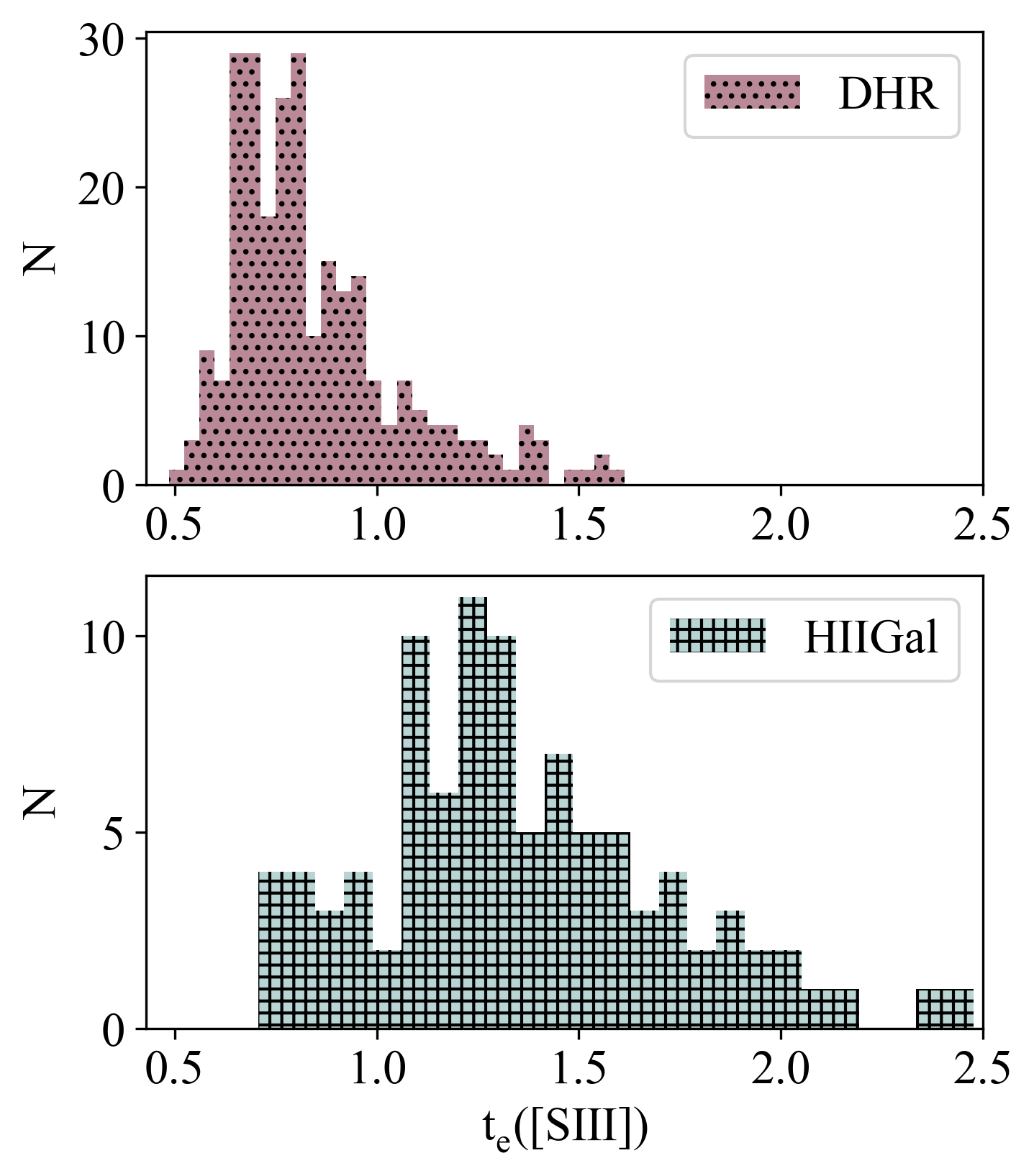}
\hspace*{0.3cm}
\includegraphics[angle=0,height=6.0cm]{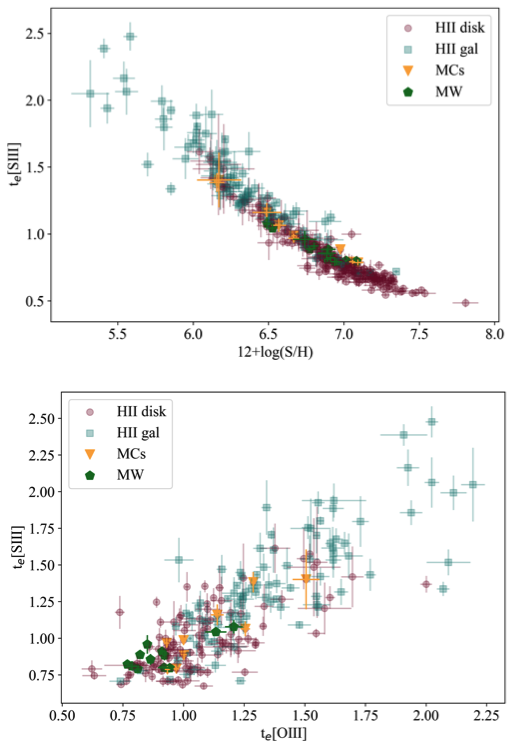}
\caption{{{\it Left:} Distribution of the derived Sulfur temperature, \TeS{III}, for the two samples studied, as labelled.}
{\it Right:} Correlation between the electron temperature, as measured by T$_e$([SIII]), and the total Sulfur abundance: 12 + log(S/H) (upper panel) . Empirical relation between [SIII] and [OIII] electron temperatures (lower panel).}
 \label{Diaz-fig3}
\end{center}
\end{figure}

Figure \ref{Diaz-fig3} (left panel) shows the \TeS{III} distribution for the two studied samples. 

The electron temperature of the ionized gas has different distributions for the two studied samples with median \TeS{III} values of  7900 K for DHR and 13000 K for  \HII Gal objects which reflects the different average metallicities of the objects in the two samples. The effect of Sulfur ions as cooling agents is evident by the good correlation shown between the total Sulfur abundance and the electron temperature as measured by \TeS{III} (Figure \ref{Diaz-fig3} (upper right panel) which is the base of the empirical calibration presented below. 
A certain number of regions also provide measurements of the [OIII] line at \ldo{4363} \AA\ so that \TeO{III} can also be derived. The relation between the two temperatures, \TeS{III} and \TeO{III} is shown in Figure \ref{Diaz-fig3} (lower right panel).

\begin{figure}  
\begin{center}
\includegraphics[angle=0,height=6.0cm]{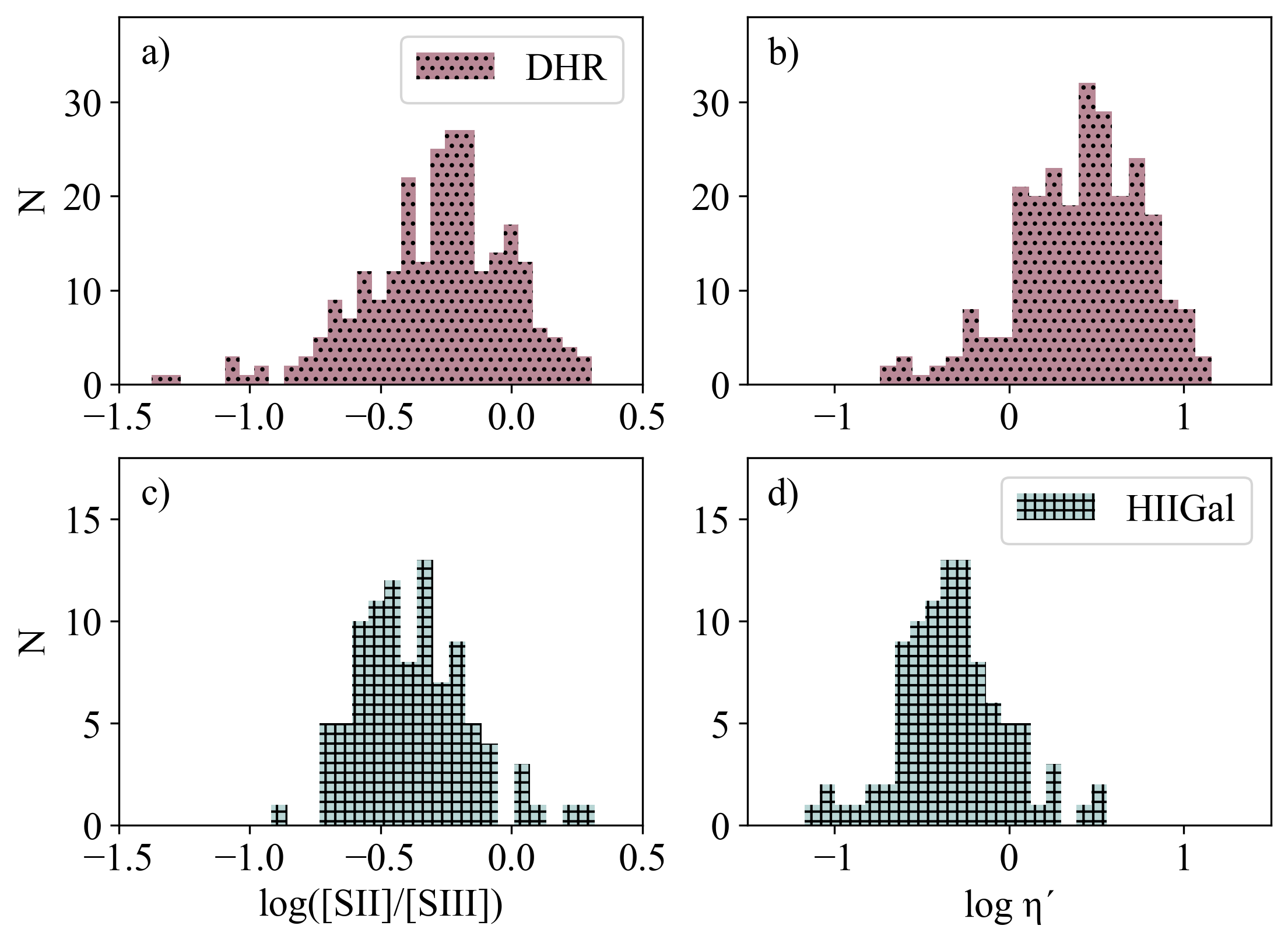}
\hspace*{0.3cm}
\caption{Distribution of the [SII]/[SIII] excitation ratio, taken as a proxy for ionization parameter (left panels), and the $\eta '$ parameter, taken as a proxy for the ionizing temperature (right panels), for DHR objects (upper panels) and \HII\ Gal objects (lower panels).} \label{Diaz-fig4}
\end{center}
\end{figure}

The two samples  show rather similar distributions regarding excitation with median values of log([SII]/[SIII]) of -0.25 and -0.38 for DHR and \HII Gal which, using the calibration by \citet{1991MNRAS.253..245D} correspond to ionization parameter values, log u,  of -2.56 and -2.34 respectively. However, they show very different distributions regarding the hardening of the ionizing radiation with median values of log $\eta$' of 0.45 for DHR and -0.36 for \HII Gal, implying equivalent effective temperatures of 38000 K and 55000 K respectively. The different distributions of ionizing temperatures seems to correspond to a real effect probably related to the nature of the ionizing clusters  with \HII Gal being characterized by their low metallicities and young ages, and DHR regions showing higher abundances (up to the solar values and maybe beyond) and, in average, being more evolved. Both high stellar metallicities and older ages would lead to lower ionizing star temperatures (see Figure \ref{Diaz-fig4}).

\subsection{Directly derived abundances}

\label{ionic abundances}

\begin{figure}  
\begin{center}
\includegraphics[angle=0,height=6.0cm]{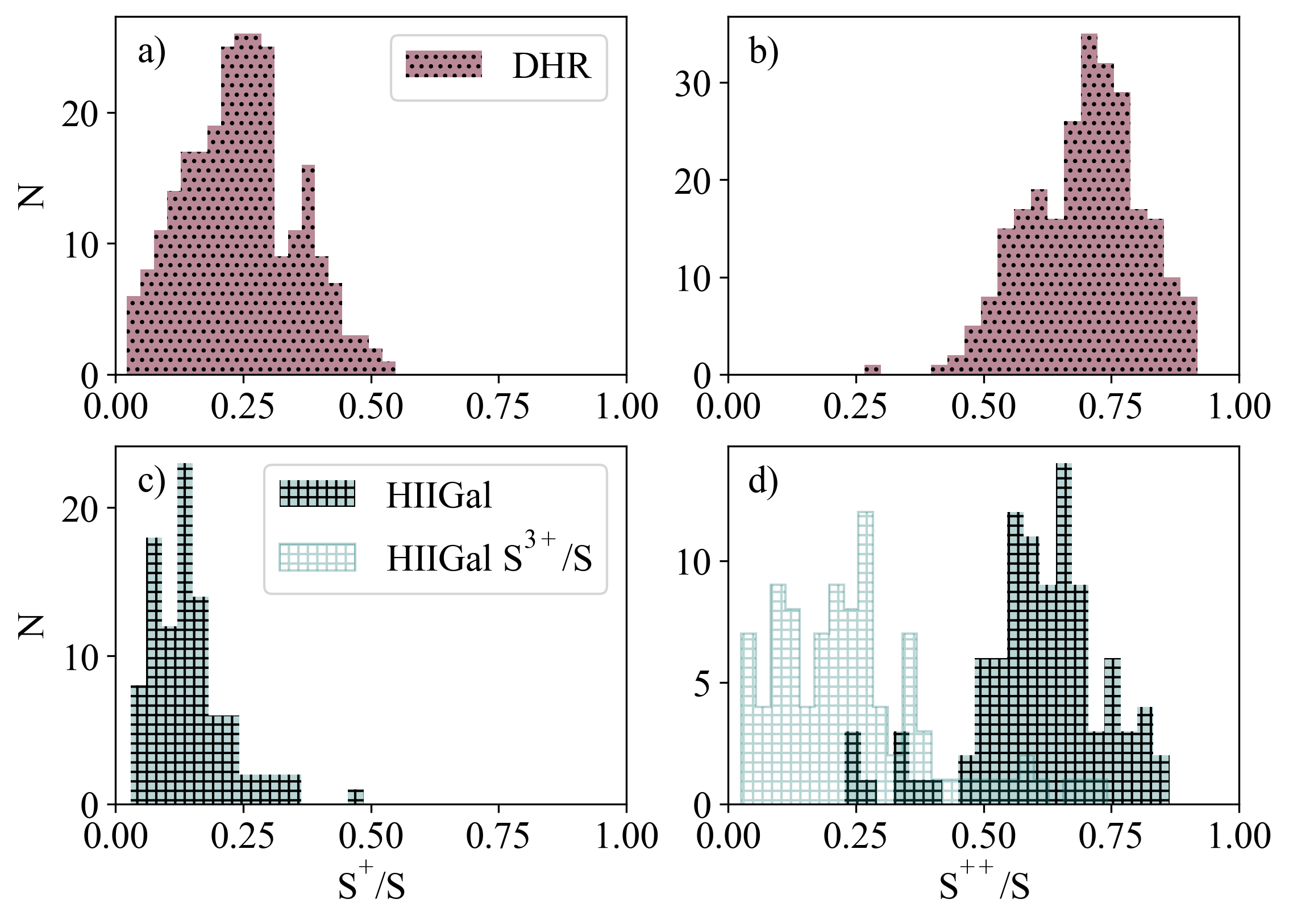}
\hspace*{0.3cm}
\caption{S$^+$ (left panels) and S$^{++}$ (right panels) contributions to the total S/H abundance for the two samples as labelled. The contribution of S$^{3+}$ in the \HII\ Gal sample is shown in the lower right panel in light green.}
\label{Diaz-fig5}
\end{center}
\end{figure}

Figure \ref{Diaz-fig5} shows the contribution of the S$^+$ and S$^{2+}$ ions to the total Sulfur abundance. For both samples, S$^{2+}$ is found to be the dominant ionization specie as can be seen in the right panel of the figure, with S$^{2+}$/S median values of about 0.7. This fact justifies the assumption of T$_e$([SIII]) as the representative electron temperature for the ionized regions. 

For DHR objects only two species, S$^+$ and S$^{++}$ coexist, and the distributions are mirrored images. In the case of \HII\ Gal objects, this is not the case since there is a certain contribution by S$^{3+}$ that can be seen in the lower right panel of the figure in light green color.

\label{Total abundances}

The S/H distributions in both samples are rather different with median values that differ by $\approx$ 0.7 dex. The highest abundance found is 
12+log(S/H)= 7.82±0.08, i.e., about 5 times the solar photospheric value \citep[12+log(S/H)$_{\odot}$ = 7.12;][]{2009ARA&A..47..481A}. The lowest Sulfur abundances, about 2\% of the solar one, are found amongst objects in the \HII\ sample..

\begin{figure}  
\begin{center}
\includegraphics[angle=0,height=4.0cm]{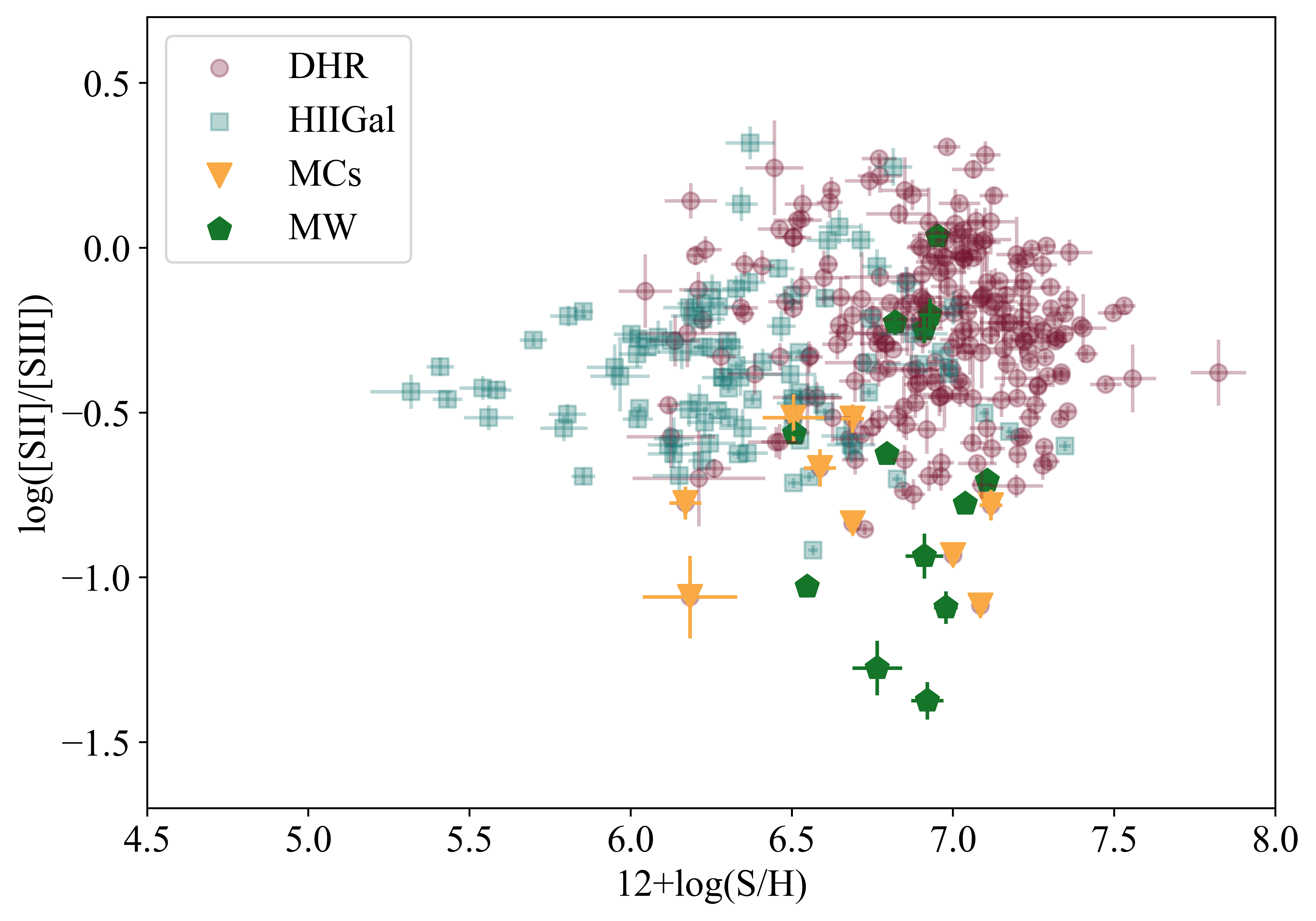}
\hspace*{0.3cm}
\includegraphics[angle=0,height=4.0cm]{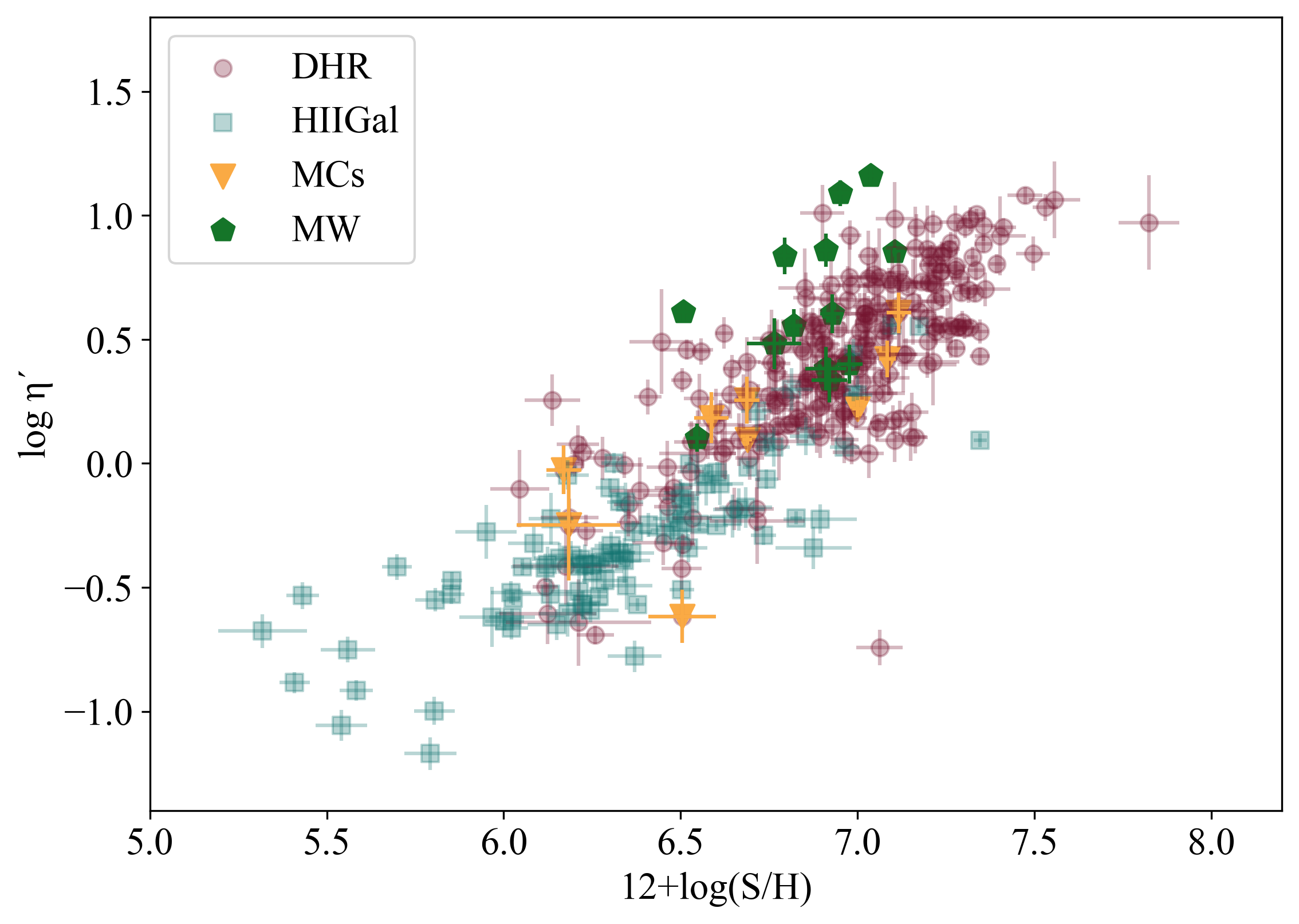}
\caption{{\it Left:} Relation between log([SII]/[SIII]), taken as a proxy for log u, and Sulfur total abundance.
{\it Right:} Relation between log($\eta$'), taken as a proxy for ionizing temperature, and Sulfur total abundance.}
 \label{Diaz-fig6}
\end{center}
\end{figure}

The relation between the functional parameters and the S/H abundance is shown in Figure \ref{Diaz-fig6}. Contrary to what is commonly assumed, no relation is found between ionization parameter and metallicity as traced by Sulfur (see the left panel of the  figure). However, there is a clear correlation between $\eta$' and metallicity (see the right panel of the  figure), showing that \HII\ galaxies are ionized by hotter sources than disk HII regions, which, as mentioned above, can be understood in terms of global metallicity (ionizing stars of higher metallicity are cooler) and age (\textbf{disc \HII regions} are probably more evolved, and their ionizing stars have lower masses and temperatures).

\begin{figure}  
\begin{center}
\includegraphics[angle=0,height=4.0cm]{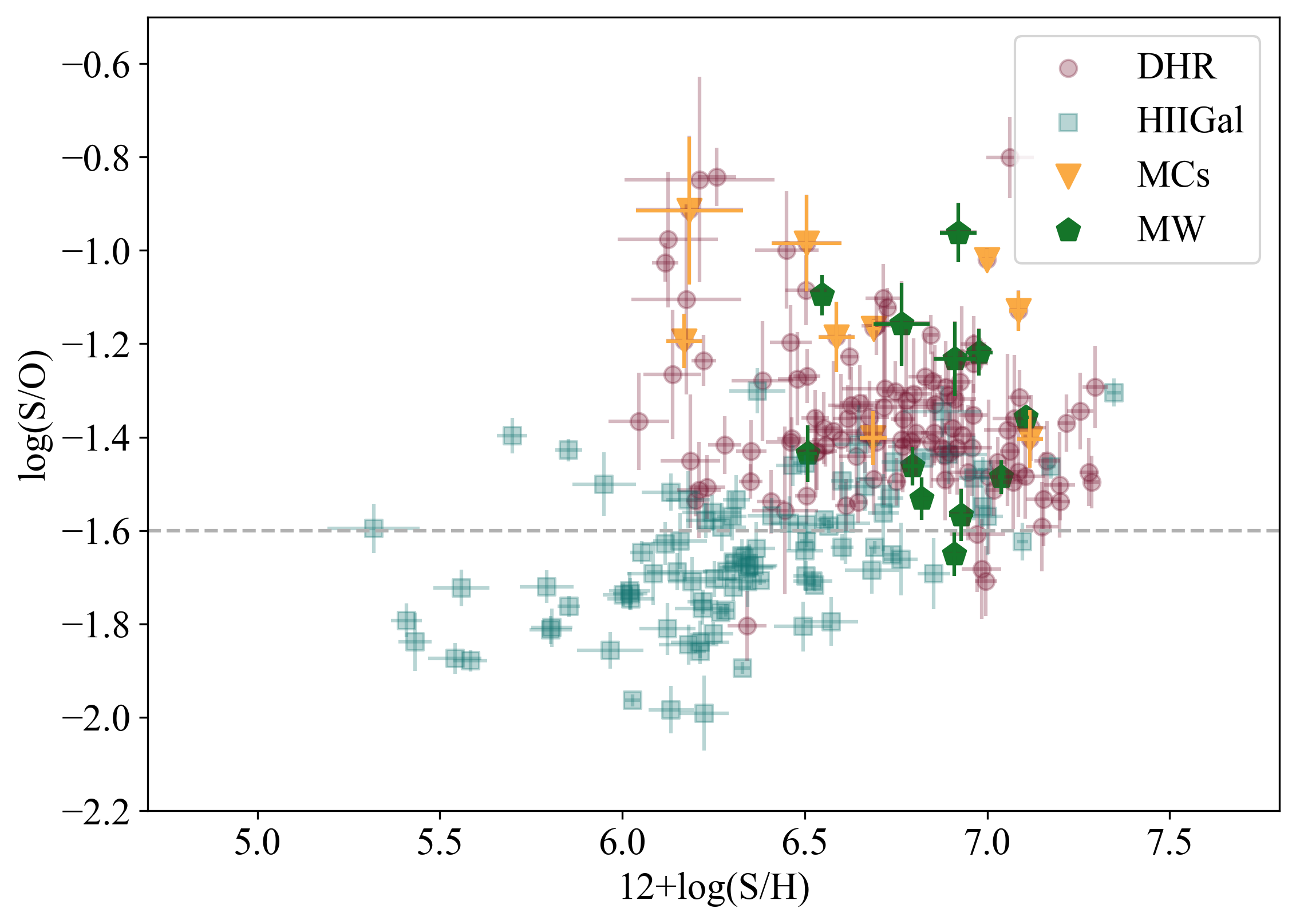}
\hspace*{0.3cm}
\caption{S/O relation against the total abundances of Sulfur.}
\label{Diaz-fig7}
\end{center}
\end{figure}

For the objects with the necessary data, it is possible to derive both the Oxygen and Sulfur abundances which allows to test the assumption of the constancy of the O/S ratio at about the solar ratio,log(S/O)$_{\odot}$ $\approx$ -1.56, if the Sulfur yield would follow that of oxygen. The derived S/O ratio can be seen in Figure \ref{Diaz-fig7} that shows: \textbf{(1)} a trend of increasing S/O with increasing S/H abundance for the \HII\ Gal sample, which could be due to the increase of oxygen depletion with increasing metallicity. On the other hand, the disk \HII\ regions seem to start at a high value of S/O and follow a decreasing trend with increasing metallicity. 

\subsection{Empirical calibrations}

The auroral lines are intrinsically weak and difficult to detect. At high metallicity, $T_{\rm{e}}$ is too low to produce any significant auroral line. However, it is possible to use the cooling properties of the ionized gas in order to produce empirical calibrations that relate emission-line intensity ratios of strong lines to the abundance of a given element. This has been done traditionally for Oxygen. One of the most popular calibrators is $R_{23}$ defined as the sum of the intensities of the [OII]\ldos{3727,29} \AA\ and [OIII]\ldos{4959,5007} \AA\ normalized to that of H$\beta$ \citep{1979MNRAS.189...95P}, that has been calibrated against directly derived abundances, which is possible only for low metallicity regions, and results from photo-ionization models in the high metallicity range. The relation obtained is two-folded, since when the O abundance increases substantially, the $T_{\rm{e}}$ in the nebula decreases and the emission lines become weaker. This does not happen for lower abundances where the cooling is dominated by free-free emission from Hydrogen and the relation has a positive slope. 

But these empirical abundance calibrations can also be done using other strong emission lines (or a combination of several), and/or  different abundance tracers. Since the strong nebular lines of Sulfur [SII] \ldos{616,31} \AA\  and [SIII] \ldos{9069,9532} \AA\  are analogous to those of Oxygen a similar calibration can be done  using the  $S_{23}$ parameter defined by S$_{23}=([SII] 6717,31+ [SIII] 9069, 9532)/H_{\beta}$ \citep{1988MNRAS.235..633V} as abundance indicator. This parameter, presents several advantages with respect to $R_{23}$: (1) the calibration remains single-valued up to, at least, solar abundances; (2) the lines are observable even at over-solar abundances; and (3) the effects of reddening are minimal. 


\begin{figure} 
\begin{center}
 \includegraphics[angle=0,height=4.0cm]{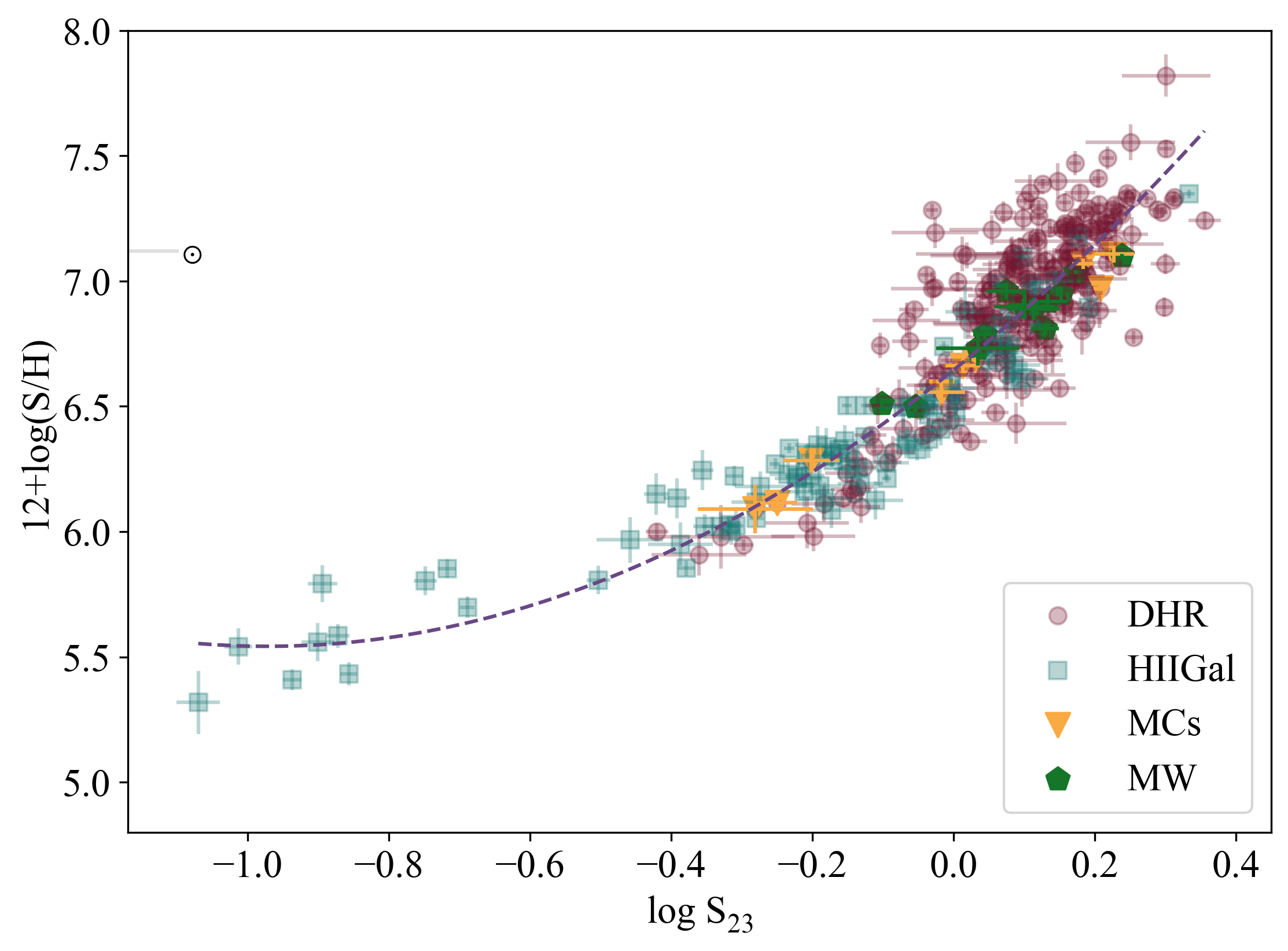}
 \hspace*{0.3cm}
 \caption{The calibration of the $S_{23}$ parameter against the total S/H abundance.}
 \label{Diaz-fig8}
 \end{center}
\end{figure}

The calibration of the $S_{23}$ parameter in terms of the total S/H abundance is shown in Figure \ref{Diaz-fig8}. A second order polynomial fit yields the relation:

\begin{equation}
\begin{split}
12+log \left(\frac{S}{H}\right)=(6.636\pm 0.011)+ (2.202\pm 0.050)\cdot log S_{23} +\\+ (1.060\pm 0.098) \cdot (log S_{23})^2
\end{split}
\label{equation_2}    
\end{equation}
\noindent with a typical deviation of 0.18. 

\begin{figure}  
\begin{center}
\includegraphics[angle=0,height=4.0cm]{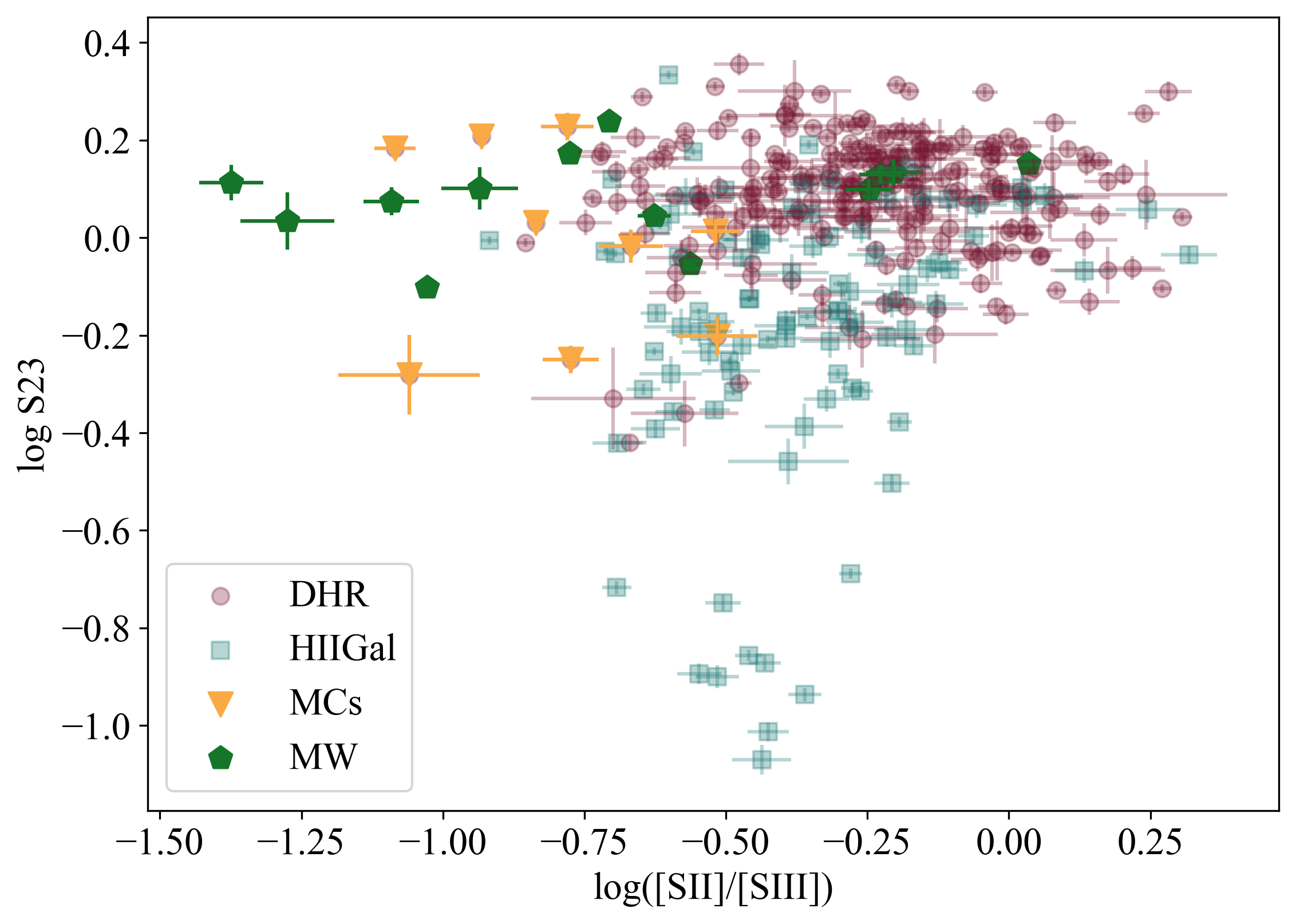}
\hspace*{0.3cm}
\includegraphics[angle=0,height=4.0cm]{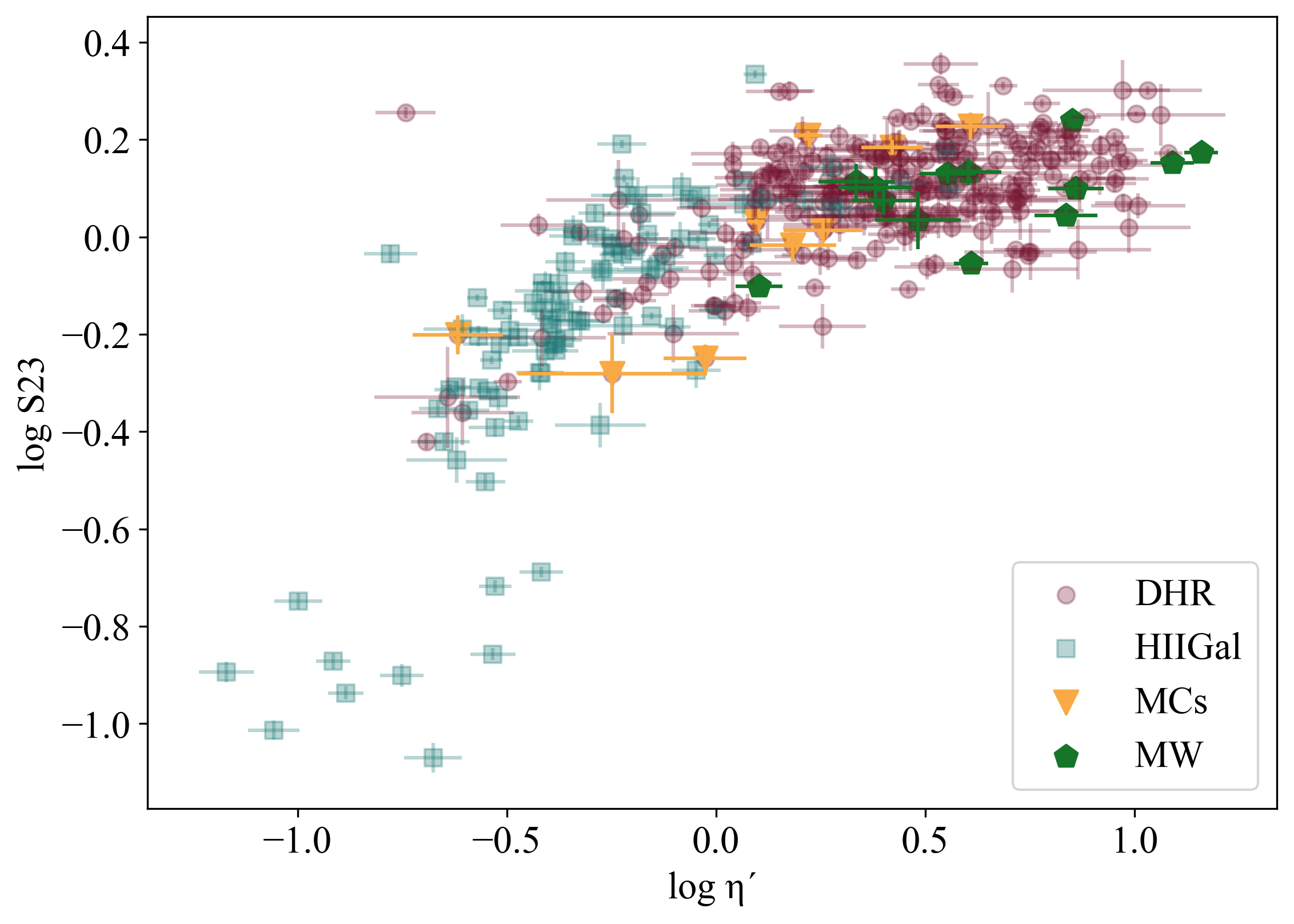}
\caption{{\it Left:} Relation between log([SII]/[SIII]), taken as a proxy for log u, and the $S_{23}$ parameter.
{\it Right:} Relation between log($\eta$', taken as a proxy for ionizing temperature, and the $S_{23}$ parameter.}
 \label{Diaz-fig9}
\end{center}
\end{figure}

The $S_{23}$ parameter is found to be independent of the ionization parameter, but shows a certain dependence of $\eta$' (see Figure \ref{Diaz-fig9}) and, as it is evident from Figure \ref{Diaz-fig8}, the dispersion in the $S_{23}$ calibration is larger than accounted for by observational errors.

\begin{figure} 
\begin{center}
 \includegraphics[angle=0,height=5.0cm]{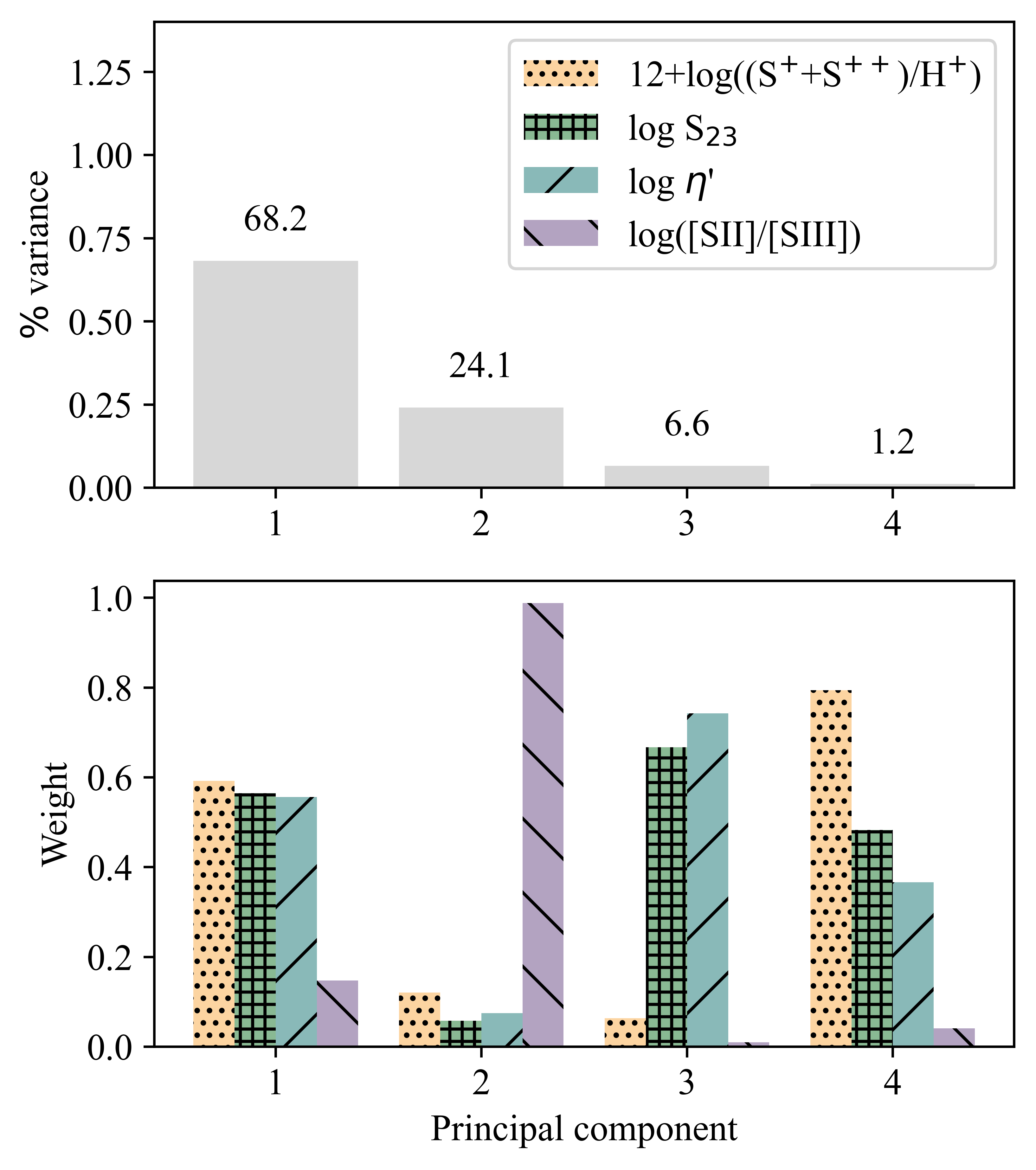}
 \hspace*{0.3cm}
 \caption{Results of the PCA showing the percentage contribution of each principal component to the total variance shown by the data set (upper panel) and the normalized relative weights of each parameter, as labelled, to each principal component (lower panel).}
  \label{Diaz-fig10}
 \end{center}
\end{figure}

To find the responsible for the intrinsic dispersion we have performed a principal component analysis. 
The obtained result indicates clearly the independence of the ionization parameter, u, from the rest of the variables and hence it has been excluded from further analysis. The relation among the other three parameters involved explains 76 \% of the variance shown by the data set, with very similar weights by each of them (see Figure \ref{Diaz-fig10}). This tendency is maintained in the analysis of the calibration residuals. According to this, it is possible to make a correction to the S/H abundance derived from the $S_{23}$ calibration to take into account the correlation of its residuals with $\eta$'. 


\section{Conclusions}
\label{conclussion}

In this work we have reviewed the methodology for the use of Sulfur to trace chemical abundances in ionized nebulae by means of spectroscopy in the red-to-near infrared wavelength range, from 600 to 980 nm. Two samples of high quality, moderate-to-high spectral resolution published data have been selected for study: (1) nebulae ionized by young massive stars located in the disks of spiral galaxies and (2) dwarf galaxies dominated by star formation bursts. The two samples show overlapping ranges of ionization parameter, however they differ widely in the value of the parameter $\eta$' indicating that \HII\ Gal objects contain hotter ionizing stellar populations.

For all the objects, we have derived the abundances of Sulfur by direct methods assuming an ionization structure composed of two zones: a low/intermediate excitation zone where S$^{++}$ and S$^+$ originate and a high ionization zone where S$^{3+}$ might be formed. For both samples, S$^{2+}$ is found to be the dominant ionization specie. In order to correct for unseen ionization stages in high excitation objects, we have used an ICF scheme based on the Ar$^{3+}$/Ar$^{3+}$ ratio which requires the detection and measurement of the [ArIV]$\lambda$ 4740 \AA , finding a good correlation between the Sulfur ICF and the parameter $\eta$'. We have applied this correlation to estimate the ICF for those objects without data on the [ArIV] line. Most DHR objects show ICFs close to unity and therefore no ICF correction needs to be applied.

The S/H abundance distributions in both samples span different ranges with median values of 12+log(S/H) of 6.27 for the \HII\  Gal sample, and 6.92 for the DHR sample with abundances reaching up to 5 times the solar photospheric value. At this high abundances, the [OIII] \ldo{4363} \AA\ auroral line is not detected and hence these objects would had been missed from the distribution had Oxygen been chosen as abundance tracer. The lowest Sulfur abundances, about 2\% of the solar value, have been found amongst \HII\ Gal objects. A good correlation exists between the Sulfur abundance and $\eta$' with low metallicity objects being ionized by stellar clusters of higher effective temperature (lower $\eta$' values).

For objects with all the necessary data, directly derived Oxygen abundances have been calculated, so that the behavior of the S/O ratio could be explored. Most  \HII\ Gal objects show S/O ratios below the solar value and a trend is found for an increased S/O ratio with increasing Sulfur abundance. On the other hand, DHR sample objects show S/O ratios larger than solar and show a tendency for lower S/O ratios at higher metallicities. More detailed work is needed, mostly in the high-metallicity regime, in order to assess the reality of these trends and understand their causes.

\acknowledgments I would like to warmly acknowledge the help of Sandra Zamora in the preparation of this review. This research has been supported by Spanish grants AYA2016-79724-C4-1-P and PID2019-107408GB-C42. 

\bibliographystyle{aaabib}
\bibliography{Diaz}

\end{document}